# The microarchitecture of a multi-threaded RISC-V compliant processing core family for IoT end-nodes

Abdallah Cheikh, Gianmarco Cerutti,
Antonio Mastrandrea, Francesco Menichelli, Mauro Olivieri[1]

[1] DIET, Sapienza University of Rome, Italy
*mauro.olivieri@uniroma1.it*

**Abstract.** Internet-of-Things end-nodes demand low power processing platforms characterized by heterogeneous dedicated units, controlled by a processor core running concurrent control threads. Such architecture scheme fits one of the main target application domain of the RISC-V instruction set. We present an open-source processing core compliant with RISC-V on the software side and with the popular Pulpino processor platform on the hardware side, while supporting interleaved multi-threading for IoT applications. The latter feature is a novel contribution in this application domain. We report details about the microarchitecture design along with performance data.

**This is the pre-print version of an article submitted to** *Lecture Notes on Electrical Engineering*, **Springer, in compliance with the Publisher's Self-archiving Policy.**

## 1  Introduction

Internet-Of-Things (IoT) end-node design demands low-power low-cost specialized processors [9,10,11]. Typical IoT end-node software runs on platforms integrating specialized units, scratchpad memories [3] and I/O units, controlled by a processing core executing multiple control threads in parallel [1], resembling the architecture depicted in Fig. 1. In such perspective, the growing interest for an extendable microprocessor instruction set has led many IoT companies to support the RISC-V open standard [2].

In this work, we present the microarchitecture solutions adopted in an open-access RISC-V compatible processor family supporting multi-threaded execution in a single core, targeting the implementation of IoT end-nodes running a limited number of concurrent control loops.

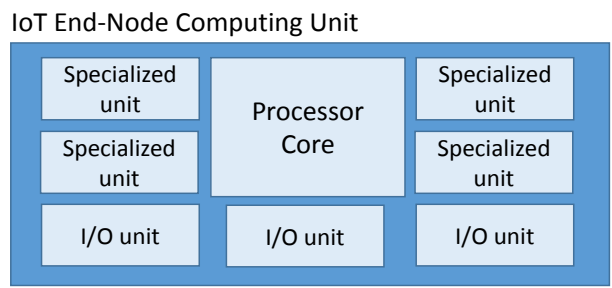

**Fig. 1.** Concept of specialized digital platform for local processing in IoT applications

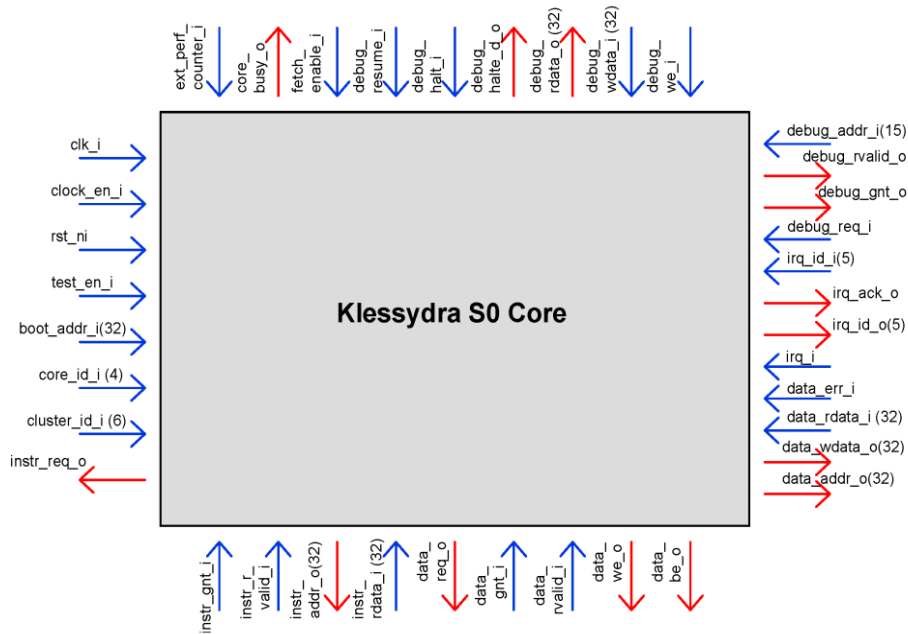

**Fig. 2.** Klessydra processing core pinout.

## 2 Architecture design specification

### 2.1 Overview

The Klessydra processing core family features full compliance with the RISC-V instruction set on the software side, and with the existing Pulpino System-on-Chip [5] on the hardware side. To date, the Klessydra family includes a single-thread core, Klessydra-S0, and a set of multi-threaded cores, Klessydra-T0, available in different implementations. All the cores of the family have been synthesized and tested on Xilinx Series 7 FPGA devices; an IC prototype design is in progress.

Klessydra cores implement the 32-bit integer RISC-V M-mode instruction set. The core interface is signal-to-signal compatible with the Pulpino microprocessor platform, and as such it is the same as RI5CY core's. The pinout of the of a typical Klessydra core contains 321 I/O signals (Fig. ).

Klessydra T0 cores implement interleaved multi-threading, i.e. at each clock cycle a new instruction is fetched from a different hardware thread. The T0 core synthesizable code is parameterized in order to have different implementations available, labelled as Klessydra-T0*BS* where *B* is the value of the Thread Pool Baseline (minimum number of active threads for stall-free pipeline operation) and *S* is the value of the Thread Pool Size (maximum number of active threads). Presently, the publicly available implementations are T022, T023, T024, T033, T034, while T012, T013, T014 are internal designs used for evaluation purposes.

## 2.2 Supported Instruction Set

Klessydra cores fully support the RV32I Base Integer Instruction Set V2.2, the M-mode privileged instruction set V1.10, and one operation of the Optional Atomic Instruction Extension RVA for thread synchronization (absent in S0 core). Table 1 reports the list of instructions implemented. Details on the operations of each instruction can be found in the RISC-V specification documents [6,7].

**Table 1.** Summary of RISC-V instructions supported by Klessydra cores

| Opcode | Funct 3/ 7/ 12 instruction operation field | Format type |
|---|---|---|
| OP_IMM | ADDI /SLTI[U] /ANDI /ORI /XORI /SLLI /SRLI /SRAI | I |
| OP | ADD /SLTI[U] /AND /OR /XOR /SLL /SRL /SRA /SUB | R |
| LUI | -- | I |
| AUIPC | -- | I |
| JAL | -- | UJ |
| JALR | -- | I |
| BRANCH | BEQ / BNE /BLT[U] /BGE[U] | SB |
| LOAD | LW /LH[U] /LB[U] | I |
| STORE | SW /SH /SB | S |
| MISC_MEM | FENCE /FENCE_I | I |
| SYSTEM | *PRIV*: "ECALL /EBREAK /MRET /WFI" /CSRRW /CSRRS /CSRRC /CSRRWI /CSRRSI /CSRRCI | I |
| AMO | AMOSWAP.W | R |

## 3 Microarchitecture building blocks

Without loss of generality, the following description refers to the Klessydra-T023 implementation, according to Fig. 3.

### 3.1 FSM_IF

The FSM_IF state machine manages instruction fetch and is able to supply one instruction per cycle when coupled with a program memory able to serve one access per cycle. Instruction are 32-bit word aligned. Compressed instruction format is not supported and no prefetch logic is present. The FSM_IF unit may stall the subsequent pipeline stages if the program memory access takes longer than one cycle.

The FSM_IF unit is not aware of multiple program counters corresponding to the different hardware threads, but simply performs the memory access with the instruction address provided by the program counter update units. The hardware thread id value (*harc*) is passed "as is" to the subsequent pipeline stages.

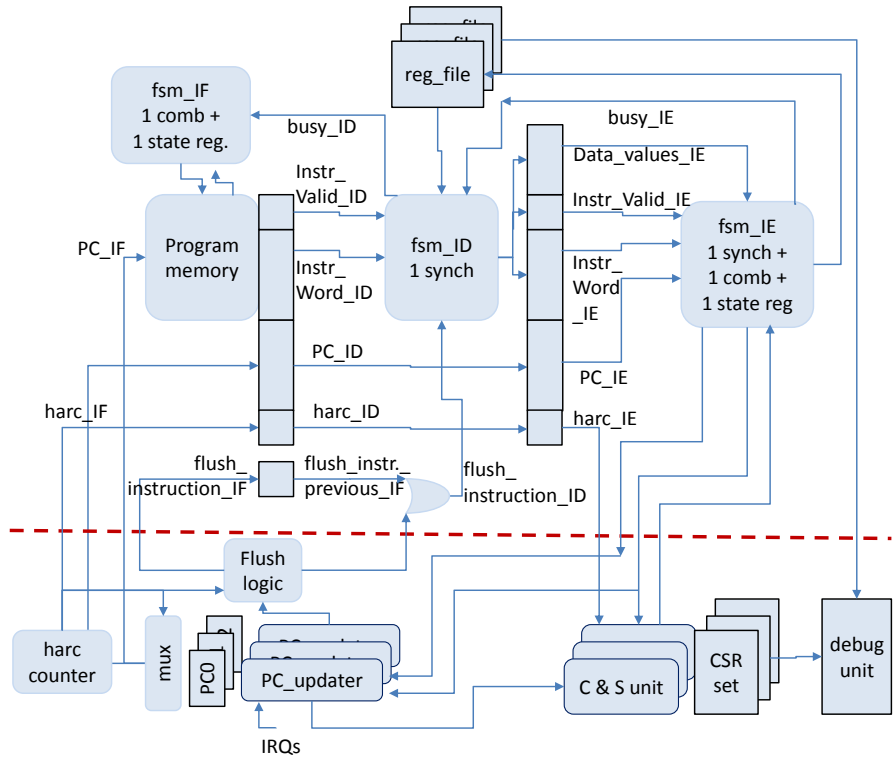

**Fig. 3.** Microarchitecture scheme (Klessydra-T023)

### 3.2 FSM_ID

The FSM_ID is a single-state synchronous unit the concurrently performs operand data fetch from the register file along with instruction operation decoding.
The decoded operation is passed to the subsequent pipeline stage in a one-hot coding format, thus generating considerable flip-flop count amplification at the advantage of largely optimizing the critical path in the architecture.

### 3.2 FSM_IE

The FSM_IE state machine handles execution/writeback of instructions. The unit can be in one of the following states:
**Sleep state**: core idle, waiting for the *fetch_enable_i* signal or an interrupt request;
**Reset State**: initial state of the core before it fetches the first instruction;
**Debug State**: core now is in debug mode, controlled by the debug inerface;
**Normal State**: The state where the processor is decoding/executing instructions;
**Data Grant:** core waiting for a grant signal to initiates new memory access;
**Data Valid Waiting State**: core waiting for data from the memory;

**CSR Instruction Wait State**: core handling CSR instruction;
**WFI Wait State**: core waiting for an interrupt to arrive (S0 core only).

The branch delay slot is managed by automatically flushing the fetched instructions *belonging to the branching thread*, when the thread pool size is insufficient to avoid branching hazards. The microarchitecture does not implement hardware interlocks for managing data dependencies. When the active threads are less than the thread pool baseline value, void threads (i.e. performing null operations) are inserted in the instruction pipeline to avoid data hazards and guarantee correct operation. Alternatively, thread data consistency may rely on classic RISC instruction scheduling to avoid hazards, as was done in [4].

### 3.3 Data Register File

Klessydra cores have 32x32-bit wide registers ranging from registers x0 to x31. Register x0 is statically bound to 0 and can only be read. Notably, the data register file is necessarily replicated on a per-thread basis.

### 3.4 Program Counter Management units

In Klessydra cores the next program counter for each hardware thread is set by the *pc_updater* units. This synchronous single-state unit updates the program counter value according to the events raised by the following signals coming from other units:
- Branch Signals: *set_branch_condition, branch_condition_pending*.
- Exception Signals: *set_except_condition, set_mret_condition*.
- Interrupt Signals: *irq_pending, served_irq*.
- Boot Signal: *rst_ni*

If none of the above are active, the unit simply increments the program counter of 4 units.

A dedicated hardware thread counter (*harc*) governs the interleaving of threads in the instruction fetch. Inactive threads (i.e. waiting for interrupt) are skipped; however when active threads are too few to guarantee data-hazard avoidance, then NOPs are automatically interleaved in the pipeline to avoid the hazards.

### 3.5 Control and Status Registers units

Each CSR unit is a state machine handling three kinds of operations on the control/status registers: CSR instruction, exception/interrupt events, and MRET instructions. Klessydra cores implement a subset of the control and status registers specified in the RISC-V privileged specification, along with some additional CSRs specifically needed for the core operations and/or for compliance with the Pulpino microprocessor platform. Such extension is composed of the MIRQ and PCER registers. The whole set of CSRs implemented in the Klessydra cores is reported in Table 2.

**Table 2.** Control and status registers supported by Klessydra cores

| Name | R/W | Description |
|---|---|---|
| MSTATUS | R/W | status register |
| MEPC | R/W | exception program counter |
| MCAUSE | R/W | trap cause |
| PCER | R/W | performance counter enable |
| MESTATUS | R/W | exception status register backup |
| MHPMCOUNTER | R/W | performance-monitoring counter |
| MHPMEVENT | R/W | performance-event selector |
| MCPUID | R | cpu description |
| MIMPID | R | implementation description |
| MHARTID | R | hardware thread integer id |
| MIP | R/W | interrupt pending type |
| MTVEC | R/W | trap-handler base address |
| MIRQ | R | ext. interrupt request number |
| MBADADDR | R/W | misaligned address value |

### 3.6 Debug Unit

The cores feature a debug unit capable of taking control of the instruction execution after an external debug request or the execution of an EBREAK instruction. The unit can operate in two modes, namely *halt mode* in which the core stops after the execution of the last fetched instruction, and *single step mode* in which the core executes one instruction at each command of the debug unit. When the debug unit is active (and the Fsm_IE is in the debug state), the data register file is accessible by the debug unit. The debug unit is compliant with the Pulpino debug interface.

## 4   Performance

Table 3 reports the throughput obtained by different implementations of the core microarchitecture. The clock cycle time refers to the present FPGA implementations on a Xilinx Series 7 device. Throughput results refer to the execution of multiple threads constituted by basic integer computational kernels. When running at baseline thread count (i.e. with active threads = Thread Pool Baseline, the four-stage-pipeline cores exhibit no pipeline stalls, while the other cores exhibit just one stall cycle per executed branch. Furthermore, according to the design expectations, the microarchitectures with higher thread pool baseline exhibit a shorter clock cycle time, as a consequence of shorter pipeline stage critical path.

Table 3. Performance results. ("-" = not applicable)

| Core module | Pipeline stages | Cycle time [ns] | Average throughput (MIPS) | | | |
|---|---|---|---|---|---|---|
| | | | 1 thread | 2 threads | 3 threads | 4 threads |
| S0   | 2 | 12.0 | 71.84 | -     | -      | -      |
| T012 | 2 | 12.7 | 67.88 | 78.74 | -      | -      |
| T022 | 3 | 8.9  | 48.43 | 96.86 | -      | -      |
| T013 | 2 | 13.9 | 62.02 | 71.94 | 71.94  | -      |
| T023 | 3 | 9.7  | 44.44 | 88.87 | 103.09 | -      |
| T033 | 4 | 7.3  | 30.58 | 59.05 | 118.09 | -      |
| T014 | 2 | 15.9 | 54.22 | 62.89 | 62.89  | 62.89  |
| T024 | 3 | 9.4  | 45.85 | 91.71 | 106.38 | 106.38 |
| T034 | 4 | 7.4  | 30.16 | 58.25 | 116.50 | 135.14 |

## 5  Conclusions

We showed the microarchitecture solutions adopted in implementing a family of processing cores compliant with RISC-V integer 32-bit instruction set and with the popular Pulpino System-on-Chip platform. Performance analysis was also reported, showing the trade-offs between different micro-architecture organizations and minimum/maximum number of threads in the target application software. Future work will be in the power efficiency analysis and in the IC implementation of the best performing micro-architecture modules. The presented work is a step towards the realization of fully open design solution database for IoT system design, from microprocessor cores to printed circuit boards and software layers.